\documentclass[aps,pra, twocolumn, amsmath, amssymb, showpacs, superscriptaddress]{revtex4-1}

\usepackage{graphicx}% Include figure files
\usepackage{bm}% bold math
\DeclareMathOperator{\Tr}{Tr}

\begin{document}

\title{Nonlinear propagation of polarized light pulses in a medium of atoms with degenerate energy levels: adiabatic approach}

\author{V. I. Yudin}
\email{viyudin@mail.ru}
\affiliation{%
Institute of Laser Physics SB RAS, pr. Akademika Lavrent'eva 13/3, Novosibirsk, 630090, Russia
}%
\affiliation{%
Novosibirsk State University, ul. Pirogova 2, Novosibirsk, 630090, Russia
}%
\affiliation{%
Novosibirsk State Technical University, pr. Karla Marksa 20, Novosibirsk, 630073, Russia
}%
\affiliation{%
Russian Quantum Center, Skolkovo, Moscow Reg., 143025, Russia
}%
\author{M. Yu. Basalaev}
\email{mbasalaev@gmail.com}
\affiliation{%
Novosibirsk State Technical University, pr. Karla Marksa 20, Novosibirsk, 630073, Russia
}%
\author{D. V. Brazhnikov}
\affiliation{%
Institute of Laser Physics SB RAS, pr. Akademika Lavrent'eva 13/3, Novosibirsk, 630090, Russia
}%
\affiliation{%
Novosibirsk State University, ul. Pirogova 2, Novosibirsk, 630090, Russia
}%
\author{A. V. Taichenachev}
\affiliation{%
Institute of Laser Physics SB RAS, pr. Akademika Lavrent'eva 13/3, Novosibirsk, 630090, Russia
}%
\affiliation{%
Novosibirsk State University, ul. Pirogova 2, Novosibirsk, 630090, Russia
}%
\affiliation{%
Russian Quantum Center, Skolkovo, Moscow Reg., 143025, Russia
}%
\date{\today}% It is always \today, today,
             %  but any date may be explicitly specified

\begin{abstract}
We develop a general method allowing one to construct the consistent theory of light pulse propagation through an atomic medium in arbitrary nonlinear regime with respect to the field strength, taking into account the light polarization, temporal (frequency) and spatial dispersions. The method is based on the reduced Maxwell equation, atomic density matrix formalism and adiabatic approximation. In order to demonstrate the efficiency of our method we investigate in detail the case of propagation of the polarization pulse under conditions of coherent population trapping in a medium of two-level atoms with degenerate energy levels. Equations describing the evolution of various field parameters are derived. It is shown that pulses of ellipticity and spatial orientation of polarization ellipse propagate with slowing. Analytical expressions for the slowing factor are obtained. A previously unknown effects of the stimulated phase modulation and the generation of a phase pilot pulse by the variation of spatial orientation of the polarization ellipse are predicted. In addition, we show that the spatial dispersion can be interpreted as ``wind effects''.
\end{abstract}


\maketitle

\section{\label{Introduction}Introduction}
Theoretical treatment of light pulse propagation through a resonant atomic medium (in the context of the classical Lorentz oscillator model) was first performed by Sommerfeld and Brillouin \cite{Sommerfeld_1907, Sommerfeld_1914, Brillouin_1960} in early 20th century. Invention of laser opened a way for experimental investigations of various resonance effects \cite{Boyd_Gauthier_2002}. The pioneer works on extreme slowing of light pulses \cite{Hau_1999, Kash_Zibrov_Lukin_1999} led to surge of interest to resonant medium with a steep dispersion in the regime of electromagnetically induced transparency (EIT) \cite{Harris,Fleischhauer_2005}. Nowadays many theoretical and experimental studies on the group velocity control in various media (atomic vapor, photonic crystals, optical fibers, semiconductor structures, microcavities, etc.) are carried out \cite{Slow_light_Book_2009}. Such attention to this field is primarily caused by promising prospects of the implementation of optical devices for information (including quantum) transmission, processing, and storage, which differ from similar electronic devices in a considerably faster operation, a higher noise immunity, and ability to preserve confidentiality of the transmitted information \cite{Phillips_Lukin_PRL_2001, Liu_Hau_Nature_2001, Matsko_Scully_2001, Lukin_colloquium_2003, Boyd_OPN_2005}. In addition, the coherent effects underlying the ``slow'' light result in giant Kerr nonlinearity \cite{Hau_1999} that is actively applied in nonlinear optics \cite{Slow_light_Book_2009, Fleischhauer_2005, Matsko_Kerr-Nonlenearity_2003}.

It should be noted that phenomena, related to the light propagation through a medium, are very diverse. A wide range of possible variations of radiation parameters and variety of media lead to a great number of physical problems. Therefore, when solving a particular problem, it is reasonable at the initial stage to use a rough physical model, which, in process of development, becomes considerably complicated and includes more and more effects. For example, the scalar model of field and the model of non-degenerate atomic energy levels are widespread in optics and spectroscopy. Although such approach to number of problems is quite adequate and qualitatively agrees with many experiments, but nevertheless it is restrictive. In reality, the light field has a vector nature, and the energy levels of atoms are degenerate with respect to the angular momentum projection. Consequently, the polarization aspect of the atom-field interaction can play a key role in description of some phenomena and processes. At the same time, this direction of research is not well developed. In overwhelming majority of works, related to the light propagation, only the amplitude modulation of wave is considered in the framework of scalar model. However, the polarization (elliptical in the general case) and its spatial orientation are degrees of freedom as well as the amplitude and the phase. Therefore, the account for the polarization aspect should lead to discovery of qualitatively new effects. Especially it concerns nonlinear regimes of the atom-field interaction, as it was repeatedly demonstrated in our works for a number of spectroscopic and laser cooling problems \cite{Prudnikov_JETP_1999, Prudnikov_JETP_2004, BTYu_JETPhL_2006, BTYu_JETPh_2009}.

In this paper, we develop a general approach to the propagation of light pulses in a resonant atomic media, which is based on the reduced Maxwell equation for the vector electromagnetic field and the atomic density matrix formalism with taking into account the Zeeman degeneracy of energy levels. The treatment is performed beyond the perturbation theory on the field amplitude. For the density matrix the adiabatic approximation is used, according to which the solution is found in the form of series in spatial and time derivatives of the slowly varying field envelope. This approach allows us to derive the nonlinear reduced Maxwell equation with account for the effects of temporal and spatial dispersions. As a concrete example, we study the propagation of elliptically polarized pulses through a gas of resonant atoms being in the coherent population trapping (CPT) state. It is known that CPT is one of the nonlinear interference effects. Its essence consists in that, under certain conditions, atoms are accumulated in a special coherent state (so-called dark state), which does not interact with the resonance field \cite{Alzetta,Agapev_UFN_1993, Arimondo_ProgInOptics_1996,Yudin_1989}. We derive an analytical expression for the group velocity of elliptically polarization pulses for all dark transitions $J \rightarrow J$ ($J$ is integer) and $J' \rightarrow J' - 1$ ($J'$  is arbitrary), which were found and described in the paper \cite{Yudin_1989}. Moreover, a novel effect of the stimulated phase modulation induced by the rotation of the polarization ellipse (i.e. the rotation of the major axes to some angle in the ellipse plane) is predicted. A particular interest is connected with the fact that at the medium entrance two light pulses, propagating at the different velocities, are generated. One of them moves at the speed of light in vacuum $c$, which we call the ``pilot'' pulse, and the other is slow pulse, which velocity can be substantially less than  $c$. It is remarkable that this nontrivial nonlinear effect occurs only in the elliptically polarized field and disappears for the cases of linear or circular polarization.

The paper is organized in the following way. In Section~\ref{Formulation_of_the_problem} we formulate the problem of slow light pulse propagation in a resonant atomic medium and present the basic equations. Section~\ref{Description_of_the_method} describes the mathematical formalism of the adiabatic approach used to solve this problem. Section~\ref{Pulses_propagation} is devoted to the exploration of elliptically polarized pulses in the medium of motionless atoms under CPT conditions. In Section~\ref{Consideration_of_the_atomic_motion} we show that the motion of atoms in a gas results in the effects of spatial dispersion. Discussion of the main results is given in Section~\ref{Conclusion}. In the appendixes \ref{Appendix:JtoJ} and \ref{Appendix:JtoJ-1} we derive the coefficients that determine the velocity of pulses for transitions  $J \rightarrow J$ ($J$ is integer) and $J' \rightarrow J' - 1$ ($J'$ is arbitrary), respectively.

%--------------------------- FORMULATION OF THE PROBLEM ------------------------------
\section{\label{Formulation_of_the_problem}Formulation of the problem}
\begin{figure}[h!]
    \includegraphics[width=0.7\linewidth]{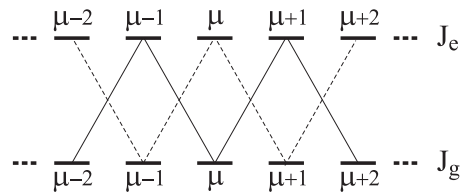}
    \caption{\label{image:two-level_atom}Scheme of light-induced transitions.}
\end{figure}

We consider the propagation of a plane electromagnetic wave through an ensemble of resonant two-level atoms with an arbitrary optical transition $J_g \rightarrow J_e$, where $J_g$ and $J_e$ are the total angular momenta of the ground and excited states, respectively. First, in order to simplify the problem and clarify the description of material, we assume that atoms to be motionless. The scheme of light-induced transitions is shown in Fig.~\ref{image:two-level_atom}, when the quantization axis is chosen along the wave vector ${\bf k}$.

Let us factor out the rapid space-time oscillations of the electric field vector ${\bf E}$ and the medium polarization vector ${\bf P}$
\begin{equation}\label{problem:E}
    \begin{aligned}
        &{\bf E} = {\bf\tilde{E}}(t,z) e^{-i\left(\omega t - k z\right)} + \text{c.c.},\\
        &{\bf P} = {\bf\tilde{P}}(t,z) e^{-i\left(\omega t - k z\right)} + \text{c.c.},
    \end{aligned}
\end{equation}
where ${\bf\tilde{E}}(t,z)$ and ${\bf\tilde{P}}(t,z)$ are the slowly varying amplitudes, $k = \omega/c$ is the wave number, $\omega$ is the frequency of electromagnetic wave, $c$ is the speed of light in free space. The slowly varying amplitude ${\bf\tilde{E}}$ satisfies the reduced Maxwell equation
\begin{equation}\label{problem:rwe}
    \left(\frac{\partial}{\partial\,z} + \frac{1}{c}\frac{\partial}{\partial\,t}\right){\bf\tilde{E}}(t,z) = 2i\pi k{\bf\tilde{P}}(t,z).
\end{equation}
The expression for polarization generated in the atomic medium by the applied electromagnetic field can be obtained from the definition of polarization vector ${\bf P}$ as the average dipole moment per unit volume:
\begin{equation}\label{problem:P_def}
    {\bf P} = N_{a}\Tr\{\hat{{\bf d}}\hat{\rho}\},
\end{equation}
where  $N_a$ is the density of atoms, $\hat{{\bf d}}$ is the dipole moment operator, $\hat{\rho}$ is the one-atomic density matrix. The quantum kinetic equation for $\hat{\rho}$ has the form
\begin{equation}\label{problem:QKE}
    \begin{aligned}
        &\frac{\partial}{\partial\,t}\hat{\rho} + \hat{\Gamma}\{\hat{\rho}\} = -\frac{i}{\hbar}[\hat{H}_0, \hat{\rho}] - \frac{i}{\hbar}[-(\hat{\bf{d}}\,\bf{E}), \hat{\rho}],\\
        &\Tr\{\hat{\rho}\} = 1,
    \end{aligned}
\end{equation}
where $\hat{H}_0$ is the Hamiltonian of unperturbed atom, $\hat{\Gamma}\left\{\hat{\rho}\right\}$ is the operator describing relaxation processes (radiative, collisional, flight, etc.).

The density matrix $\hat{\rho}$ can be separated in four matrix blocks
\begin{equation}\label{problem:rho_blocks}
    \hat{\rho} = \hat{\rho}_{gg} + \hat{\rho}_{ee} + \hat{\rho}_{eg} + \hat{\rho}_{ge},
\end{equation}
which in the basis of Zeeman states $\left|J, \mu\right\rangle$ have the form
\begin{equation}\label{problem:rho_elements}
    \hat{\rho}_{ab} =\sum_{\mu_a, \mu_b}\rho_{\mu_a, \mu_b}^{a\,b}\left|J_a, \mu_a\right\rangle\left\langle J_b, \mu_b\right|, \quad (a, b) = (e, g),
\end{equation}
where $\rho_{\mu_a, \mu_b}^{a\,b}$ are the matrix elements. Hereinafter the index ``$g$'' is connected with the ground energy level of an atom, and the index ``$e$'' is connected with the excited one. The diagonal matrix blocks $\hat{\rho}_{gg}$ and $\hat{\rho}_{ee}$ describe the population of atomic states and the low-frequency (Zeeman) coherences, the off-diagonal matrix blocks $\hat{\rho}_{eg}$ and $\hat{\rho}_{ge}$ ($\hat{\rho}_{ge}=\hat{\rho}_{eg}^{\dag}$) correspond to the optical coherences.

By factoring in the optical coherences the fast space-time oscillations at the field frequency
\begin{equation}\label{problem:change_opt_coh}
    \begin{aligned}
        &\hat{\rho}_{eg} = \hat{\tilde{\rho}}_{eg}e^{-i(\omega\,t - k\,z)},\\
        &\hat{\rho}_{ge} = \hat{\tilde{\rho}}_{ge}e^{i(\omega\,t - k\,z)}
    \end{aligned}
\end{equation}
and using the rotating-wave approximation, we obtain from the Eq.~(\ref{problem:QKE}) the following system of generalized optical Bloch equations (GOBE):
\begin{eqnarray}
   &&\left(\frac{\partial}{\partial\,t} - i\delta\right)\hat{\tilde{\rho}}_{eg} +  \hat{\Gamma}_{eg}\{\hat{\tilde{\rho}}\} = \frac{i}{\hbar}\left(\hat{V}\hat{\rho}_{gg} - \hat{\rho}_{ee}\hat{V}\right),
    \label{problem:GOBE_rho_eg}\\
   &&\left(\frac{\partial}{\partial\,t} + i\delta\right)\hat{\tilde{\rho}}_{ge} + \hat{\Gamma}_{ge}\{\hat{\tilde{\rho}}\} = \frac{i}{\hbar}\left(\hat{V}^{\dagger}\hat{\rho}_{ee} - \hat{\rho}_{gg}\hat{V}^{\dagger}\right),
    \label{problem:GOBE_rho_ge}\\
    &&\frac{\partial \hat{\rho}_{ee}}{\partial\,t} + \hat{\Gamma}_{ee}\{\hat{\tilde{\rho}}\} = \frac{i}{\hbar}\left(\hat{V}\hat{\tilde{\rho}}_{ge} - \hat{\tilde{\rho}}_{eg}\hat{V}^{\dagger}\right),
    \label{problem:GOBE_rho_ee}\\
    &&\frac{\partial \hat{\rho}_{gg}}{\partial\,t} + \hat{\Gamma}_{gg}\{\hat{\tilde{\rho}}\} = \frac{i}{\hbar}\left(\hat{V}^{\dagger}\hat{\tilde{\rho}}_{eg} - \hat{\tilde{\rho}}_{ge}\hat{V}\right),
    \label{problem:GOBE_rho_gg}\\
    &&\Tr\left\{\hat{\rho}_{ee}\right\} + \Tr\left\{\hat{\rho}_{gg}\right\} = 1.
    \label{problem:GOBE_norm_cond}
\end{eqnarray}
Here we introduce the following notations
\begin{equation}\label{problem:rho_tilde}
    \hat{\tilde{\rho}} = \hat{\rho}_{gg} + \hat{\rho}_{ee} + \hat{\tilde{\rho}}_{eg} + \hat{\tilde{\rho}}_{ge}
\end{equation}
and
\begin{eqnarray}\label{problem:V_D_def}
&& \hat{V} = \hat{{\bf d}}_{eg}{\bf\tilde{E}}=\sum_{q=0,\pm 1}(\hat{{\bf d}}_{eg})_{q}({\bf\tilde{E}})^q,\; (\hat{{\bf d}}_{eg})_{q} = \left\langle J_e\right||\hat{d}\left||J_g\right\rangle \hat{D}_q,\nonumber \\
&& \hat{D}_q  = \sum_{\mu_g, \mu_e}\left|J_e, \mu_e\right\rangle\,C_{J_g\mu_g1q}^{J_e\mu_e}\left\langle J_g, \mu_g\right|,
\end{eqnarray}
where $\left\langle J_e\right||\hat{d}\left||J_g\right\rangle$ is the reduced matrix element of the dipole moment, $C_{J_g\mu_g1q}^{J_e\mu_e}$ are the Clebsch-Gordan coefficients, $\delta = (\omega - \omega_{eg})$ is the detuning of the field frequency $\omega$ from the transition frequency $\omega_{eg}$.

An exact description of the light pulse propagation requires a self-consistent solution of the equations for the density matrix  (\ref{problem:GOBE_rho_eg})-(\ref{problem:GOBE_norm_cond}) and reduced Maxwell equation (\ref{problem:rwe}) for the slowly varying field amplitude. Such a solution can be found using numerical methods, while with increasing of the transition angular momenta calculations become more complicated and time-taking. However, the theoretical analysis can be significantly simplified and clarified in the framework of adiabatic approximation of the medium response. In particular, using the method described below, some problems can be solved analytically.

%------------------------------- DESCRIPTION OF THE METHOD ------------------------
\section{\label{Description_of_the_method}Description of the method}
This section outlines the mathematical formalism of the method. The restriction on the minimum pulse duration, under which the field amplitude changes slightly at a time of establishment the stationary state (the adiabatic approximation of the medium response) allows us to split the problem into two successive steps. At the first stage the density matrix and the polarization vector ${\bf\tilde{P}}$ are calculated. The second stage is solution of the Eq.~(\ref{problem:rwe}) for the slowly varying amplitude ${\bf\tilde{E}}$, taking into account  the dependence of the polarization vector on the field vector ${\bf\tilde{P}}({\bf\tilde{E}})$ obtained on the first step.

The system of GOBE (\ref{problem:GOBE_rho_eg})-(\ref{problem:GOBE_norm_cond}) can be represented in the operator form:
\begin{equation}\label{method:L}
    \frac{\partial}{\partial t}\hat{\tilde{\rho}} = \hat{L}({\bf \tilde{E}})\{\hat{\tilde{\rho}}\}
\end{equation}
where $\hat{L}({\bf \tilde{E}})\{\hat{\tilde{\rho}}\}$ is the linear functional operator, depending on the vector ${\bf \tilde{E}}$. We will find a solution of the Eq.~(\ref{method:L}), using a perturbation theory, in the form of a series on the time derivatives (of the slowly varying field envelope):
\begin{equation}\label{method:expansion_rho}
    \hat{\tilde{\rho}} = \hat{\tilde{\rho}}^{(0)} + \hat{\tilde{\rho}}^{(1)} + \hat{\tilde{\rho}}^{(2)} + \ldots, \quad \hat{\tilde{\rho}}^{(k)} \sim \frac{\partial^k}{\partial\,t^k},
\end{equation}
i.e. the operator $\partial/\partial t$ is a ``smallness parameter'' of the expansion. Substituting $\hat{\tilde{\rho}}$ in the form of series (\ref{method:expansion_rho}) into Eq.~(\ref{method:L}) and equating terms of the same order, we obtain the following system of equations
\begin{eqnarray}
    \label{method:rho_0}
    \hat{L}\{\hat{\tilde{\rho}}^{(0)}\} &=& 0,\\
    \label{method:rho_1}
    \hat{L}\{\hat{\tilde{\rho}}^{(1)}\} &=& \frac{\partial}{\partial t}\,\hat{\tilde{\rho}}^{(0)},\\
    \label{method:rho_2}
    \hat{L}\{\hat{\tilde{\rho}}^{(2)}\} &=& \frac{\partial}{\partial t}\,\hat{\tilde{\rho}}^{(1)},\\
    &\ldots&\nonumber\\
    \label{method:rho_n}
    \hat{L}\{\hat{\tilde{\rho}}^{(n)}\} &=&  \frac{\partial}{\partial t}\,\hat{\tilde{\rho}}^{(n-1)},
\end{eqnarray}
supplemented by the normalization condition
\begin{equation}\label{method:cond_save_popul}
    \Tr\{\hat{\tilde{\rho}}^{(0)}\} = 1, \quad \Tr\{\hat{\tilde{\rho}}^{(n)}\} = 0 \quad (n = 1,\,2,\,\ldots).
\end{equation}
The matrix $\hat{\tilde{\rho}}^{(0)}$, which depends on the time via the slowly varying field amplitude ${\bf \tilde{E}}(t)$, can be fiound from Eqs.~(\ref{method:rho_0}) and (\ref{method:cond_save_popul}). Then, substituting  $\hat{\tilde{\rho}}^{(0)}$ in the right-hand side of Eq.~(\ref{method:rho_1}), we calculate $\hat{\tilde{\rho}}^{(1)}$, which will be of the first order in $\partial {\bf \tilde{E}}(t)/\partial t$. The higher terms of the series (\ref{method:expansion_rho}) are found in a similar way. The equations for them can be written in the form of recurrent formula (\ref{method:rho_n}).

From (\ref{problem:E}), (\ref{problem:P_def}), (\ref{problem:rho_blocks}) and (\ref{problem:change_opt_coh}) we obtain the expression for the slowly varying amplitude  ${\bf\tilde{P}}$:
\begin{equation}\label{method:Pt_def_from_micro-param}
    {\bf\tilde{P}}(t,z) = N_{a}\Tr\{\hat{{\bf d}}_{ge}\,\hat{\tilde{\rho}}_{eg}\},
\end{equation}
Then, substituting $\hat{\tilde{\rho}}_{eg}$ in the last expression in the form of the series (\ref{method:expansion_rho}), we find
\begin{equation}\label{method:expansion_rho_eg}
    {\bf\tilde{P}} = N_{a}\Tr\{\hat{{\bf d}}_{ge}\,\hat{\tilde{\rho}}^{(0)}_{eg}\} + N_{a}\Tr\{\hat{{\bf d}}_{ge}\,\hat{\tilde{\rho}}^{(1)}_{eg}\} + \ldots ~,
\end{equation}
which can be represented in the following form:
\begin{equation}\label{method:Pt_series}
    {\bf\tilde{P}}(t,z) ={\bf\tilde{P}}^{(0)} + {\bf\tilde{P}}^{(1)} + {\bf\tilde{P}}^{(2)} +  \ldots ~,
\end{equation}
where the first two terms can be written as following expressions:
\begin{align}
    &{\bf\tilde{P}}^{(0)} = \hat{M}^{(0)}_{1}{\bf \tilde{E}} + \hat{M}^{(0)}_{2}{\bf \tilde{E}}^{*},\nonumber\\
    &{\bf\tilde{P}}^{(1)} = \hat{M}^{(1)}_{1}\frac{\partial {\bf \tilde{E}}}{\partial t} + \hat{M}^{(1)}_{2}\frac{\partial {\bf \tilde{E}}^{*}}{\partial t}\,.
    \label{method:Pt_matrix_expansion}
\end{align}
In the general case, the matrices $\hat{M}^{(0)}_{1,2}$ and $\hat{M}^{(1)}_{1,2}$ depend on the electric field vector ${\bf \tilde{E}}$ and its complex conjugate ${\bf \tilde{E}}^{*}$.

The first term ${\bf\tilde{P}}^{(0)}$ of the expansion (\ref{method:Pt_series}) corresponds to the steady-state solution for $\hat{\tilde{\rho}}^{(0)}$ (see Eq.~(\ref{method:rho_0})), the second term  ${\bf\tilde{P}}^{(1)}$  takes into account the dispersion of the medium and determines the propagation velocity of the slow field envelope, the third term ${\bf\tilde{P}}^{(2)}$ is responsible for the group velocity dispersion and the pulse shape distortion. The remaining terms of the series (\ref{method:Pt_series}) describe higher order dispersion. In this work, we focus only on the first two terms ${\bf\tilde{P}}^{(0)}$ and ${\bf\tilde{P}}^{(1)}$.

It is obviously that the described above the mathematical procedure for the density matrix and the medium polarization vector (as a series expansion on ${\partial^k}/{\partial\,t^k}$) can be easily extended to a configuration with an arbitrary number of energy levels and any number of light waves.

%--------------------- PULSES PROPAGATION -------------------------
\section{\label{Pulses_propagation}Pulse propagation}
In this section we investigate the case of the resonant light pulses propagation in a medium with a dark transition $J_g \rightarrow J_e$. The feature of this transition consists in that there is so-called dark state, which is a coherent superposition of the ground state sublevels and does not interact with light field: $\hat{V}|dark\rangle = 0$ \cite{Yudin_1989}.

In the absence of relaxation in the ground state and taking into account only the radiative relaxation of the exited level $J_e$, we find that the vectors  ${\bf\tilde{P}}^{(0)}$ and ${\bf\tilde{P}}^{(1)}$ have the form
\begin{equation}\label{pulses:P_vec}
    \begin{aligned}
        &{\bf\tilde{P}}^{(0)} = 0,\\
        &{\bf\tilde{P}}^{(1)}(t,z) = i\hbar N_a \hat{M}\,\frac{\partial}{\partial\,t}{\bf\tilde{E}},
    \end{aligned}
\end{equation}
where the elements of the matrix $\hat{M}$ are given by formulas (\ref{J_J:M}) and (\ref{J_J-1:M}) for the transitions  $J \rightarrow J$ ($J$ is integer) and $J' \rightarrow J' - 1$ ($J$ is arbitrary), respectively (see Appendices \ref{Appendix:JtoJ} and \ref{Appendix:JtoJ-1}). Comparing (\ref{pulses:P_vec}) with the general formulae (\ref{method:Pt_matrix_expansion}), we see that $\hat{M}^{(0)}_{1,2} = \hat{M}^{(1)}_{2} = 0$, $\hat{M}^{(1)}_{1} \equiv \hat{M}$, that is conditioned by the peculiarity of dark transitions.

\begin{figure}[h!]
    \includegraphics[width=0.6\linewidth]{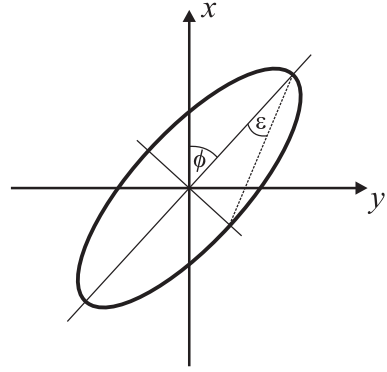}
    \caption{\label{image:ellipse_polarization} Parametrization of the polarization vector {\bf e} (\ref{pulses:e_def}).}
\end{figure}

Let us use the real amplitude $A$, the common phase $\alpha$ and the unit complex polarization vector ${\bf e}$ to parametrize the vector ${\bf\tilde{E}}$:
\begin{equation}\label{pulses:def_Et}
    {\bf\tilde{E}} = A e^{i\alpha}\,{\bf e}, \quad {\bf e} = \sum_{q}e^{\,q}{\bf e}_{q},
\end{equation}
where $e^{\,\mu}$ are the contravariant components of vector ${\bf e}$. In the cyclic basis $\left\{{\bf e}_0 = {\bf e}_z, {\bf e}_{\pm1} = \mp({\bf e}_x \pm i{\bf e}_y)/\sqrt{2}\right\}$ the vector ${\bf e}$ has the form
\begin{equation}\label{pulses:e_def}
    {\bf e} = -\cos{(\varepsilon - \pi/4)}e^{-i\phi}{\bf e}_{+1} - \sin{(\varepsilon - \pi/4)}e^{i\phi}{\bf e}_{-1},
\end{equation}
where $\varepsilon$ is the ellipticity angle and $\phi$ is the orientation angle of the light polarization ellipse in space, as shown in Fig.~\ref{image:ellipse_polarization}. Thus, a set of the real-valued parameters $\{A, \alpha, \varepsilon, \phi\}$ fully describes state of the light field, it has a transparent physical meaning and represents the natural parametrization of the electric field vector.  Each of these parameters should be considered as an independent function of the time $t$ and the spatial coordinate $z$.

It can be shown that vectors ${\bf e}$ and ${\bf e}_0$ are the eigenvectors of matrix $\hat{M}$ with zero eigenvalues $m = 0$ è $m_0 = 0$, respectively:
\begin{equation}\label{pulses:m_m0}
    \hat{M}{\bf e} = 0, \quad \hat{M}{\bf e}_0 = 0.
\end{equation}
The third eigenvector ${\bf e}_{\bot}$
\begin{equation}\label{pulses:M e_bot=m e_bot}
    \hat{M}{\bf e}_{\bot} = m_{\bot}{\bf e}_{\bot}
\end{equation}
is found as the orthogonal to the first two (${\bf e}_{\bot} = [{\bf e} \times {\bf e}_0]$):
\begin{equation}\label{pulses:e_{bot}_def}
    {\bf e}_{\bot} = \sin{(\varepsilon - \pi/4)}e^{-i\phi}{\bf e}_{+1} - \cos{(\varepsilon - \pi/4)}e^{i\phi}{\bf e}_{-1}
\end{equation}
and can be expressed as follows
\begin{equation}\label{pulses:e_bot}
    {\bf e}_{\bot} = \frac{\partial\,{\bf e}}{\partial\,\varepsilon}.
\end{equation}
Using the representation of the vector ${\bf \tilde{E}}$ in the form (\ref{pulses:def_Et}) we calculate its derivatives with respect to space and time variables ($\eta = z,\,t$):
\begin{equation}\label{pulses:dEtdeta}
    \frac{\partial {\bf\tilde{E}}}{\partial \,\eta} = \left(\frac{\partial A}{\partial \eta} + iA\frac{\partial \alpha}{\partial \eta}\right)e^{i\alpha}{\bf e} + A e^{i\alpha}\left(\frac{\partial \varepsilon}{\partial \eta}\,\frac{\partial \,{\bf e}}{\partial \varepsilon} + \frac{\partial \phi}{\partial \eta}\,\frac{\partial \,{\bf e}}{\partial \phi}\right).
\end{equation}
Since the vectors ${\bf e}$ and ${\bf e}_{\bot}$ are mutually orthogonal $\left({\bf e}^{*}\cdot{\bf e}_{\bot} = 0\right)$ and the vector $\partial \,{\bf e}/\partial \phi$ lies in the same plane, then $\partial \,{\bf e}/\partial \phi$ can be represented as a linear combination of ${\bf e}$  and ${\bf e}_{\bot}$:
\begin{equation}\label{pulses:relation_e_et}
    \frac{\partial \,{\bf e}}{\partial \phi} = - i\sin{\left(2\varepsilon\right)}{\bf e} - i\cos{\left(2\varepsilon\right)}{\bf e}_{\bot}.
\end{equation}
Then
\begin{eqnarray}\label{pulses:dEtdeta}
    \frac{\partial {\bf\tilde{E}}}{\partial \,\eta} &=& \left(\frac{\partial A}{\partial \eta} + iA\frac{\partial \alpha}{\partial \eta} - iA\sin{\left(2\varepsilon\right)}\frac{\partial \phi}{\partial\eta}\right)e^{i\alpha}{\bf e}+\nonumber\\
    &+& A\left(\frac{\partial \varepsilon}{\partial \eta} - i\cos{\left(2\varepsilon\right)}\frac{\partial \phi}{\partial \eta} \right)e^{i\alpha}{\bf e}_{\bot}.
\end{eqnarray}
From (\ref{pulses:P_vec}), (\ref{pulses:dEtdeta}) ($\eta = t$) and taking into account (\ref{pulses:m_m0}), (\ref{pulses:M e_bot=m e_bot}) we obtain the expression for the medium polarization vector
\begin{equation}\label{J_J:P}
    {\bf\tilde{P}}(t,z) = i\hbar N_{a}m_{\bot} A\,e^{i\alpha} \left[\frac{\partial\varepsilon}{\partial\,t} - i\cos{(2\varepsilon)}\frac{\partial\phi}{\partial\,t}\right]{\bf e}_{\bot}.
\end{equation}
Substituting the last expression in the reduced Maxwell equation (\ref{problem:rwe}), after some mathematical transformations, we obtain the full system of equations, which describes the space-time evolution of real-valued field parameters
\begin{align}
  \label{pulses:rwe_A0}
    &\left(\frac{\partial}{\partial z} + \frac{1}{c}\frac{\partial}{\partial t}\right)A = 0,\\
  \label{pulses:rwe_epsilon}
    &\left(\frac{\partial}{\partial z}  + \frac{1}{c}\left[1 + s(\varepsilon, A)\right]\frac{\partial}{\partial t}\right)\varepsilon = 0,\\
  \label{pulses:rwe_phi}
    &\left(\frac{\partial}{\partial z}  + \frac{1}{c}\left[1 + s(\varepsilon, A)\right]\frac{\partial}{\partial t}\right) \phi = 0,\\
  \label{pulses:rwe_alpha0}
    &\left(\frac{\partial}{\partial z} + \frac{1}{c}\frac{\partial}{\partial t}\right) \alpha - \sin{(2\varepsilon)}\left(\frac{\partial}{\partial z} + \frac{1}{c}\frac{\partial}{\partial t}\right) \phi = 0.
\end{align}
Here
\begin{equation}\label{pulses:s}
    s(\varepsilon, A) = 2\pi N_a \hbar\,\omega\,m_{\bot}(\varepsilon, A)
\end{equation}
is the coefficient determining the slowing-down of pulses. It depends on the field amplitude $A$ and ellipticity $\varepsilon$, and is independent of the angle $\phi$ and phase $\alpha$. Derivation of the expression for the quantity $m_\bot$ in analytical form is sufficiently cumbersome and  is given in Appendices: formula  (\ref{J_J:m_bot_final}) for the transition  $J \rightarrow J$ and formula (\ref{J_J-1:m_bot_final}) for the transition $J' \rightarrow J' - 1$.

Notable is the fact that the differential equations (\ref{pulses:rwe_A0})-(\ref{pulses:rwe_alpha0}) on the field parameters are ``decoupled'' and form a hierarchy, that is for each subsequent solution of the equation is required to find solutions of the previous equations. Generally, there are four types of waves corresponding to the modulation of the amplitude $A$, ellipticity  $\varepsilon$, orientation angle of the polarization ellipse  $\phi$, and phase  $\alpha$. From the first equation (\ref{pulses:rwe_A0}) it is seen that the amplitude modulation travels at the velocity of light in free space and it is of no interest. The second (\ref{pulses:rwe_epsilon}) and third (\ref{pulses:rwe_phi}) equations are more remarkable, they imply that the pulses of ellipticity $\varepsilon(t, z)$ and angle $\phi(t, z)$ propagate with slowing. At the same time, in contrast to the ellipticity wave, the differential equation for $\phi$ is linear. Consequently, if the amplitude and the ellipticity are stationary $\left(A = const,\,\varepsilon = const\right)$ then the pulse of angle propagates without distortions at group velocity $v_g = c/(1 + s)$, i.e. $\phi = \phi\left(t - z/v_g\right)$. Time delay, in this case, is easily retuned in a wide range by modifying the ellipticity or amplitude. For some applications, it can be more convenient to modulate just the orientation angle of the polarization ellipse (``to waggle'' the axis of ellipse) rather than the ellipticity. If in this case a polarizer is placed at the output of the cell, the modulation of the angle is transformed to intensity modulation.

Note that a close problem was solved by Zelenskii and Mironov \cite{Zelencky_Mironov_2002}. They derived the equations describing the evolution of the field in terms of the complex polarization parameter $q$ and intensity. They also found numerically the slowing factor  $s$ for some dark transitions in the formalism of the state amplitudes. In our parametrization, the quantity $q$ from \cite{Zelencky_Mironov_2002} has the form
\begin{equation}\label{pulses:parameter_q}
    q = e^{+1}/e^{-1} = -\tan{(\varepsilon + \pi/4)}\,e^{-i2\phi},
\end{equation}
i.e. the complex parameter $q$ depends on two real parameters $\varepsilon$ and $\phi$. From (\ref{pulses:rwe_epsilon}) and (\ref{pulses:rwe_phi}) it can be easily shown that any functional of the form $F(\varepsilon)G(\phi)$ (and, in particular, $q$) satisfies the equation
\begin{equation}
    \left(\frac{\partial}{\partial z} + \frac{1}{c}\left[1 + s(\varepsilon, A)\right]\frac{\partial}{\partial t}\right)F(\varepsilon)G(\phi) = 0.
\end{equation}
At the same time, it is not obvious from \cite{Zelencky_Mironov_2002} that $\arg{(q)}$ according to its physical meaning is related to the angle $\phi$ because such interpretation and the respective analysis are absent in the paper \cite{Zelencky_Mironov_2002}.

However, the most interesting effects is the anomalous behavior of the phase $\alpha(t,z)$ (see Eq.~(\ref{pulses:rwe_alpha0})), whose space-time variations previously have not been considered. The essence of this effect is that two induced pulses, fast and slow, are generated under the rotation of the polarization ellipse (i.e., under the modulation of the angle $\phi$) even despite the absence of phase perturbations at the boundary. For the mathematical description of this effect, it is necessary to solve the hierarchy of Eqs.~(\ref{pulses:rwe_A0})-(\ref{pulses:rwe_alpha0}) exactly in the order they are written. The simplest case is that of the constant amplitude and ellipticity. Then, assuming that the atomic medium starts at  $z = 0$, we find from Eq.~(\ref{pulses:rwe_phi}) the solution for the variation of the angle $\phi(t,z) = \phi\left(t - z/v_g\right)$, where $v_g = c/[1 + s(\varepsilon,A)]$ is group velocity, and from Eq.~(\ref{pulses:rwe_alpha0}) we obtain the stimulated solution for the phase $\alpha$
\begin{equation}\label{pulses:epsA0-const_alpha0}
        \alpha(t,z) = \alpha_0 + \sin{(2\varepsilon)}\left[\phi\left(t - z/v_g\right) - \phi\left(t - z/c\right)\right],
\end{equation}
which satisfies the boundary condition $\alpha(t, z=0) = \alpha_0$. As is seen from Eq.~(\ref{pulses:epsA0-const_alpha0}), two phase pulses simultaneously emerge under the rotation of the light polarization ellipse. One of them $\propto\phi\left(t - z/c\right)$ is a ``pilot'' pulse and it moves at the speed of light in free space  $c$. The second pulse $\propto\phi\left(t - z/v_g\right)$ moves synchronously with the angle modulation. Remarkably, this effect can be observed only in elliptically polarized light and disappears for linear or circular  polarization. In the case of linear polarization ($\varepsilon = 0$), it follows directly from Eqs.~(\ref{pulses:rwe_alpha0}), (\ref{pulses:epsA0-const_alpha0}). For the circular polarization ($\varepsilon = \pm \pi/4$) the notion of spatial orientation angle $\phi$ is meaningless. Indeed, putting in expression (\ref{pulses:e_def}) $\varepsilon = \pm \pi/4$, the electric field vector can be written as ${\bf\tilde{E}} = A \exp(i\alpha^{\prime}){\bf e}_{\pm1}$ and instead the system (\ref{pulses:rwe_A0})-(\ref{pulses:rwe_alpha0}) we have only two equations of the form $\partial X/\partial z + 1/c \, \partial X/\partial t = 0$ for the amplitude $A$ and phase $\alpha^{\prime}$.

Note that the propagation dynamics of the polarization pulses in vacuum  ($N_a = 0$) is described by the system of Eqs.~(\ref{pulses:rwe_A0})-(\ref{pulses:rwe_alpha0}) at $s(\varepsilon, A) = 0$. Although the formal relation between the phase $\alpha$ and angle $\phi$ (see Eq.~(\ref{pulses:rwe_alpha0})) remains the same as for the atomic medium with CPT, however, taking into account equation (\ref{pulses:rwe_phi}) at $s = 0$, we find that the phase $\alpha$ in vacuum satisfies the homogeneous differential equation
\begin{equation}\label{pulses:eq_alpha_vacuum}
    \left(\frac{\partial}{\partial z} + \frac{1}{c}\frac{\partial}{\partial t}\right)\alpha = 0
\end{equation}
and is independent of the angle $\phi$. Thus, the effect of stimulated excitation of phase pulses under the rotation of the polarization ellipse is absent in vacuum.

%--------------------------------- ATOMIC GAS -------------------------------------
\section{\label{Consideration_of_the_atomic_motion}Influence of the atomic motion (spatial dispersion)}
Effects caused by the spatial dispersion represent a great interest. The idea of using a strong spatial dispersion of refractive index for the effective control of the light pulses velocity in gas of   $\Lambda$-atoms was proposed in \cite{Kocharovskaya_PRL_2001}. Therefore, it arises the natural question: how the spatial dispersion effects are described in our approach.

The spatial nonlocality of the medium response due to the atomic motion, can be taken into consideration by replacing $\partial/\partial t$ $\Rightarrow$ $(\partial/\partial t + u_{z}\,\partial/\partial z)$ in the equation (\ref{problem:QKE}):
\begin{equation}\label{gas:QKE}
    \left(\frac{\partial}{\partial\,t} + u_{z}\frac{\partial}{\partial\,z}\right)\hat{\rho}+ \hat{\Gamma}\{\hat{\rho}\} = -\frac{i}{\hbar}\left[\hat{H}_0, \hat{\rho}\right] - \frac{i}{\hbar}\left[-(\hat{\bf{d}}\,\bf{E}), \hat{\rho}\right],
\end{equation}
where the Wigner density matrix $\hat{\rho}(t,z,u_z)$ depends on the atomic velocity $u_{z}$. Assuming now $(\partial/\partial t + u_{z}\,\partial/\partial z)$ as small quantity of the density matrix expansion (\ref{method:expansion_rho}), instead of the system (\ref{method:rho_0})-(\ref{method:rho_n}) we have the following equations:
\begin{eqnarray}
    \label{gas:rho_0}
    \hat{L}\{\hat{\tilde{\rho}}^{(0)}\} &=& 0,\\
    \label{gas:rho_1}
    \hat{L}\{\hat{\tilde{\rho}}^{(1)}\} &=& \left(\frac{\partial}{\partial t} + u_{z}\frac{\partial}{\partial z}\right)\hat{\tilde{\rho}}^{(0)},\\
    &\ldots&\nonumber\\
    \label{gas:rho_n}
    \hat{L}\{\hat{\tilde{\rho}}^{(n)}\} &=&  \left(\frac{\partial}{\partial t} + u_{z}\frac{\partial}{\partial z}\right)\hat{\tilde{\rho}}^{(n-1)}.
\end{eqnarray}
In this case the operator $\hat{L}\{...\}$ arises from Eq.~(\ref{method:L}) by the replacing $\delta\Rightarrow(\delta -ku_z)$.

Repeating the similar described above calculating procedure for the vectors ${\bf\tilde{P}}^{(0)}$ and ${\bf\tilde{P}}^{(1)}$ we obtain
\begin{eqnarray}\label{gas:P_v}
    &&{\bf\tilde{P}}(t, z, u_z) = \hat{M}^{(0)}_{1}(u_z) {\bf \tilde{E}} + \hat{M}^{(0)}_{2}(u_z) {\bf \tilde{E}}^{*} +
    \nonumber\\
    &&\hat{M}^{(1)}_{1}(u_z)\Big(\frac{\partial {\bf\tilde{E}}}{\partial\,t} + u_z \frac{\partial {\bf\tilde{E}}}{\partial\,z}\Big)
    + \hat{M}^{(1)}_{2}(u_z)\Big(\frac{\partial {\bf\tilde{E}}^{*}}{\partial\,t} + u_z \frac{\partial {\bf\tilde{E}}^{*}}{\partial\,z}\Big).\nonumber\\
\end{eqnarray}
In Eq.~(\ref{problem:rwe}) for the slowly varying field amplitude the polarization vector ${\bf\tilde{P}}(t, z, u_z)$ must be averaged over atomic velocity:
\begin{equation}\label{gas:rwe}
    \left(\frac{\partial}{\partial\,z} + \frac{1}{c}\frac{\partial}{\partial\,t}\right){\bf\tilde{E}}(t,z) = 2i\pi k\langle{\bf\tilde{P}}\rangle_{u_z},
\end{equation}
where
\begin{eqnarray}\label{gas:P_average}
    &&\langle{\bf\tilde{P}}\rangle_{u_z} = \langle\hat{M}^{(0)}_{1}\rangle_{u_z}{\bf \tilde{E}} + \langle\hat{M}^{(0)}_{2}\rangle_{u_z}{\bf \tilde{E}}^{*} + \langle\hat{M}^{(1)}_{1}\rangle_{u_z}\frac{\partial {\bf\tilde{E}}}{\partial\,t} +
    \nonumber\\
    &&\langle u_z\hat{M}^{(1)}_{1}\rangle_{u_z}\frac{\partial {\bf\tilde{E}}}{\partial\,z} + \langle\hat{M}^{(1)}_{2}\rangle_{u_z}\frac{\partial {\bf\tilde{E}}^{*}}{\partial\,t} + \langle u_z\hat{M}^{(1)}_{2}\rangle_{u_z}\frac{\partial {\bf\tilde{E}}^{*}}{\partial\,z}.
    \nonumber\\
\end{eqnarray}
The symbol $\langle Q\rangle_{u_z}$ designates the averaging
\begin{equation}
    \langle Q\rangle_{u_z}= \int_{-\infty}^{+\infty} Q(u_z) f(u_z)\,du_z\nonumber
\end{equation}
over atomic velocity distribution $f(u_z)$.
Thus, exactly the terms, which contain the spatial derivative $\partial /\partial z$ in Eq.(\ref{gas:P_average}), are responsible for the spatial dispersion effects.

Let us continue the consideration of the dark transitions $J \rightarrow J$ ($J$ is integer) and $J' \rightarrow J' - 1$ ($J'$ is arbitrary) from the previous section. Because the matrices $\hat{M}^{(1)}_{1} =  \hat{M}$ and $\hat{M}^{(1)}_{2} = \hat{M}^{(0)}_{1,2} = 0$ are independent of the atomic speed $u_z$, the polarization vector ${\bf\tilde{P}}(t,z)$ is given by the expression
\begin{equation}\label{gas:P_vec_dark}
    {\bf\tilde{P}}(t,z) = i\hbar N_a \hat{M}\frac{\partial {\bf\tilde{E}}}{\partial\,t} + i\hbar N_a\bar{u}  \hat{M}\frac{\partial {\bf\tilde{E}}}{\partial\,z},
\end{equation}
where $\bar{u}=\langle u_z\rangle_{u_z}$ is the atomic average velocity along the wave vector $k$ (i.e. along the $z$-axis).

Thus, instead of Eq.~(\ref{pulses:rwe_A0})-(\ref{pulses:rwe_alpha0}) the evolution of the field parameters is described now by the following equations
\begin{align}
  \label{gas:rwe_A0}
    &\frac{\partial A}{\partial\,z} + \frac{1}{c}\frac{\partial A}{\partial\,t} = 0,\\
  \label{gas:rwe_epsilon}
    &\left[1 + \frac{\bar{u}s(\varepsilon,A)}{c}\right]\frac{\partial \varepsilon}{\partial\,z} + \frac{1}{c}\left[1 + s(\varepsilon,A)\right]\frac{\partial \varepsilon}{\partial\,t} = 0,\\
  \label{gas:rwe_phi}
    &\left[1 + \frac{\bar{u}s(\varepsilon,A)}{c}\right]\frac{\partial \phi}{\partial\,z} + \frac{1}{c}\left[1 + s(\varepsilon,A)\right]\frac{\partial \phi}{\partial\,t} = 0,\\
  \label{gas:rwe_alpha0}
    &\frac{\partial \alpha}{\partial\,z} + \frac{1}{c}\frac{\partial \alpha}{\partial\,t} - \sin{(2\varepsilon)}\left(\frac{\partial \phi}{\partial\,z} + \frac{1}{c}\frac{\partial \phi}{\partial\, t}\right) = 0.
\end{align}
To transform the Eqs.~(\ref{gas:rwe_epsilon}) and (\ref{gas:rwe_phi}) to the canonical form we divide them by coefficient $[1 + \bar{u}s(\varepsilon,A)/c]$ in front of the space derivative:
\begin{eqnarray}
    \label{gas:rwe_epsilon_canonical}
        \frac{\partial \varepsilon}{\partial\,z} + \frac{1 + s(\varepsilon,A)}{c + \bar{u}s(\varepsilon,A)}\frac{\partial \varepsilon}{\partial\,t} = 0,\\
    \label{gas:rwe_phi_canonical}
        \frac{\partial \phi}{\partial\,z} + \frac{1 + s(\varepsilon,A)}{c + \bar{u}s(\varepsilon,A)}\frac{\partial \phi}{\partial\,t} = 0.
\end{eqnarray}
From these equations we see that the group velocity of the pulses propagating in atomic gas is given by
\begin{equation}\label{gas:vg}
    v_{g} = \frac{c + \bar{u}s(\varepsilon,A)}{1 + s(\varepsilon,A)} \approx \frac{c}{s(\varepsilon,A)} +\bar{u} \quad (\text{if}\; s(\varepsilon,A)\gg 1).
\end{equation}
Let us consider two cases. First case, atoms in gas obey  maxwellian distribution with peak at $u_{z} = 0$:
\begin{equation}\label{gas:U_velocity_distribution}
    f(u_z) = \frac{1}{u_{T}\sqrt{\pi}}\exp{\left(-\frac{u_{z}^2}{u_{T}^2}\right)},
\end{equation}
where $u_{T} = \sqrt{2 k_B T/m_{a}}$  is the most probable thermal velocity. Then the atomic average velocity vanishes $\bar{u} = 0$. Therefore, the system of equations for the field parameters (\ref{gas:rwe_A0})-(\ref{gas:rwe_alpha0}) takes the same form as for motionless atoms (\ref{pulses:rwe_A0})-(\ref{pulses:rwe_alpha0}). Other situation, the velocity distribution has a selected speed $u_{0}$, i.e. it is shifted from zero (for example, the case of pulse propagation in a flow of atoms):
\begin{equation}\label{gas:U_jet}
    \widetilde{f}(u_z) = \frac{1}{u_{T}\sqrt{\pi}}\exp{\left(-\frac{(u_{z} - u_{0})^2}{u_{T}^2}\right)}.
\end{equation}
In this case, the average velocity of the atoms is different from zero $\bar{u} \neq 0$. Consequently, rate of atomic flow $u_0$  is another parameter that allows to control the speed of the polarization pulses. Analyzing the formula (\ref{gas:vg}), same the work \cite{Kocharovskaya_PRL_2001} for $\Lambda$-system it can be identified three regimes:
\begin{equation}
    a)\,0 < v_{g} < c, \quad b)\,v_{g} < 0, \quad c)\,v_{g} = 0.
\end{equation}
The regime $b$ occurs if atoms move opposite to the light pulses and $\bar{u} < - c/s(\varepsilon,A)$, the regime $c$ corresponds to the condition $\bar{u} = - c/s(\varepsilon,A)$, in all other cases take place the regime $a$. Note that in contrast to work \cite{Kocharovskaya_PRL_2001} our results (including the formula (\ref{gas:vg})) are obtained outside of the linear approximation on the weak field.

Performed above analysis demonstrates that the spatial dispersion effects in the atomic gas are closely related to the movement of atoms and these effects can be interpreted as the effects of spatial transfer of the medium polarization. Indeed, when light pulse moving opposite to the atomic flow ($\bar{u}<0$), then the ``wind slowing-down'' of the pulse occurs. And vice versa, by co-directional movement of light and gas flow ($\bar{u}>0$) the pulse group velocity rises relative to the flow absence ($\bar{u} = 0$), i.e. the ``wind speeding-up'' of the light pulses occurs.

Note, that in the considered case of the dark transitions the matrices $\hat{M}^{(1)}_{1}$ and $\hat{M}^{(1)}_{2}$ are independent of the atomic speed $u_z$ and the one-photon detuning $\delta$, because we investigate the idealized case of a plane wave and take into account only the radiative relaxation. However, in the more realistic case of transversally limited light beams and/or considering the collisional decoherence in the ground state, the matrices $\hat{M}^{(1)}_{1}$ and $\hat{M}^{(1)}_{2}$ will depend on both the velocity of atoms and the one-photon detuning. Changing the one-photon detuning $\delta$ we pick out a group of atoms satisfying condition $\delta = k u_z$, which resonantly interact with the laser field. As a result, $\langle u_z\hat{M}^{(1)}_{1,2}\rangle_{u_z} \neq 0$  in the general case even for a symmetric velocity distribution (including the maxwellian distribution (\ref{gas:U_velocity_distribution})) at $\delta \neq 0$. Thereby, mentioned above the ``wind effects'' will be determined by the speed of the resonance atomic group $(\delta = k u_z)$.

%--------------------------------- CONCLUSION -------------------------------------
\section{\label{Conclusion}Conclusion}
In the present paper, we developed the general method, which allows us in the adiabatic approximation sequentially to describe a process of the optical pulses propagation in {\em arbitrary nonlinear} regime with respect to the field strength. This approach enables a natural way to take into account both the effects of temporal dispersion due to the persistence of the medium response and the effects of spatial dispersion due to the atomic motion.

As an example, we have investigated in detail the case of polarization pulses of monochromatic field propagating through the medium of two-level atoms with a dark transition  $J_g \rightarrow J_e$ in the approximation of only radiative relaxation. The system of differential equations (\ref{gas:rwe_A0})-(\ref{gas:rwe_alpha0}), completely describing the space-time dynamics of all real-valued field parameters (the amplitude $A$, the common phase $\alpha$, the ellipticity angle $\varepsilon$ and the spatial orientation angle of field polarization ellipse $\phi$), was derived. Proper choice of the field parametrization resulted in that these equations have a simple mathematical form and a clear physical interpretation. Nontrivial fact is that the equations (\ref{gas:rwe_A0})-(\ref{gas:rwe_alpha0}) form the certain hierarchy and require solution in exactly that sequence as they are written, since the solution of each next equation of the hierarchy require solutions of all overlying equations. From the analysis of the system (\ref{gas:rwe_A0})-(\ref{gas:rwe_alpha0}) it follows that the amplitude pulse $A(t - z/c)$ travels at speed of light in free space, whereas the pulses of ellipticity $\varepsilon(t,z)$ and orientation of the polarization ellipse propagate with slowing. For the coefficients defining velocity of this pulses we obtained analytical expressions for transitions $J_g = J \rightarrow J_e = J$ ($J$ is integer) with one dark state and for transitions $J_g = J' \rightarrow J_e = J' - 1$ ($J'$ is arbitrary) with two dark states. However, the most interesting result is related to the specific behavior of the field phase $\alpha$. In accordance with Eq.~(\ref{gas:rwe_alpha0}) the change of spatial orientation of polarization ellipse yields the stimulated modulation of the phase  $\alpha$.  At the same time two phase pulses are exited at the atomic medium boundary: ``pilot'' pulse moving at the speed of light in vacuum and slow pulse propagating with delay. Note that predicted pilot pulse effect takes place only in an elliptically polarized field.

The presented  results provide a better understanding of the polarized radiation propagation in nonlinear atomic media and can be applied in the field of optical communications. Emphasize, that the described method can be generalized to the case of an arbitrary number of the field frequency components and atoms with any energy levels structure.

The work has been supported by the Ministry of Education and Science of the Russian Federation in the frame of the Program ``Scientific and scientific-pedagogical personnel of innovative Russia'' (the Contract no. 16.740.11.0466 and the Agreement no. 8387), by RFBR (grants nos. 12-02-00454, 12-02-00403, 11-02-00775, 11-02-01240), by the Russian Academy of Sciences and Presidium of Siberian Branch of Russian Academy of Sciences. The young scientists D.V. Brazhnikov and M.Yu. Basalaev have been also supported by the Presidential Grant MK-3372.2912.2 and the RFBR Grant 12-02-31208.

%--------------------- SLOWING FACTORS FOR THE J -> J TRANSITIONS ----------------------
\appendix
\section{\label{Appendix:JtoJ}Transition $J_g = J \rightarrow J_e = J$ ($J$ is integer)}
Taking into consideration only the radiative relaxation $\gamma$ of the exited level $J_e$, we obtain from (\ref{problem:GOBE_rho_eg})-(\ref{problem:GOBE_norm_cond}) the following standard system of GOBE \cite{TTY_PRA_2004}:
\begin{align}
\label{appendix:GOBE_rho_eg}
    &\left(\frac{\partial}{\partial\,t} + \frac{\gamma}{2} - i\delta\right)\hat{\tilde{\rho}}_{eg} = \frac{i}{\hbar}\left(\hat{V}\hat{\rho}_{gg} - \hat{\rho}_{ee}\hat{V}\right),
    \\
\label{appendix:GOBE_rho_ge}
    &\left(\frac{\partial}{\partial\,t} + \frac{\gamma}{2} + i\delta\right)\hat{\tilde{\rho}}_{ge} = \frac{i}{\hbar}\left(\hat{V}^{\dagger}\hat{\rho}_{ee} - \hat{\rho}_{gg}\hat{V}^{\dagger}\right),
    \\
 \label{appendix:GOBE_rho_ee}
    &\left(\frac{\partial}{\partial\,t} + \gamma\right)\hat{\rho}_{ee} = \frac{i}{\hbar}\left(\hat{V}\hat{\tilde{\rho}}_{ge} - \hat{\tilde{\rho}}_{eg}\hat{V}^{\dagger}\right),~
   \\
 \label{appendix:GOBE_rho_gg}
   &\frac{\partial}{\partial\,t}\hat{\rho}_{gg} - \gamma\sum_{q=0,\pm 1}{\hat{D}_{q}^{\dagger}\hat{\rho}_{ee}{\hat{D}_{q}}} = \frac{i}{\hbar}\left(\hat{V}^{\dagger}\hat{\tilde{\rho}}_{eg} - \hat{\tilde{\rho}}_{ge}\hat{V}\right),
    \\
\label{appendix:GOBE_norm_cond}
    &\Tr\left\{\hat{\rho}_{ee}\right\} + \Tr\left\{\hat{\rho}_{gg}\right\} = 1.
\end{align}
Let us consider the case of transitions $J_g = J \rightarrow J_e = J$ ($J$ is integer), for which in accordance with work \cite{Yudin_1989} in monochromatic elliptically polarized field there is only one dark state  $|\psi_{0}^{(g)}\rangle$ satisfying the equation
\begin{equation}\label{J_J:dark_state}
    \hat{V}|\psi_{0}^{(g)}\rangle = 0.
\end{equation}
Following \cite{TTY_PRA_2004}, we introduce the basis of eigenfunctions of operators $\hat{V}^{\dagger}\hat{V}$ and $\hat{V}\hat{V}^{\dagger}$:
\begin{equation}\label{J_J:basis_eigenstates}
    \begin{aligned}
        \hat{V}\hat{V}^{\dagger}|\psi_{i}^{(e)}\rangle = |\lambda_i|^2|\psi_{i}^{(e)}\rangle,\\
        \hat{V}^{\dagger}\hat{V}|\psi_{j}^{(g)}\rangle = |\lambda_j|^2|\psi_{j}^{(g)}\rangle.
    \end{aligned}
\end{equation}
States $|\psi_{i}^{(e)}\rangle$ and $|\psi_{j}^{(g)}\rangle$ form so-called ``natural'' basis of ground and excited levels, respectively. In this basis, the interaction operator $\hat{V}$ is a diagonal matrix
\begin{equation}\label{J_J:V_def_diag}
    \hat{V} = \sum_{j}\lambda_{j}|\psi_{j}^{(e)}\rangle\langle\psi_{j}^{(g)}|.
\end{equation}
The dark state $|\psi_0^{(g)}\rangle$ corresponds to  $\lambda_0 = 0$:
\begin{equation}\label{J_J:V_psi_0}
    \hat{V}|\psi_{0}^{(g)}\rangle = \lambda_{0}|\psi_{0}^{(e)}\rangle = 0.
\end{equation}
Zeroth-order approximation $\hat{\tilde{\rho}}^{(0)}$ in the expansion (\ref{method:expansion_rho}) is defined only by the dark state $|\psi_0^{(g)}\rangle$:
\begin{equation}\label{J_J:rho_gg_0}
    \hat{\rho}_{gg}^{(0)} = |\psi_{0}^{(g)}\rangle\langle\psi_{0}^{(g)}|, \quad \hat{\rho}_{ee}^{(0)} = 0, \quad \hat{\tilde{\rho}}_{eg\,(ge)}^{(0)} = 0.
\end{equation}
Using the representation of the density matrix in natural basis (\ref{J_J:basis_eigenstates}), from the system of Eqs.~(\ref{appendix:GOBE_rho_eg}) -- (\ref{appendix:GOBE_norm_cond}) taking into account the zero-order approximation (\ref{J_J:rho_gg_0}) it can be shown that $\rho_{ee}^{(1)} = 0$, à $\hat{\tilde{\rho}}_{eg}^{(1)}$ and $\hat{\tilde{\rho}}_{ge}^{(1)}$ have the form
\begin{equation}\label{J_J:rho_eg-rho_ge}
    \hat{\tilde{\rho}}_{eg}^{(1)} = \sum_{j}\rho^{eg}_{j\,0}|\psi_{j}^{(e)}\rangle\langle\psi_{0}^{(g)}|,\quad \hat{\tilde{\rho}}_{ge}^{(1)} = \sum_{j}\rho^{ge}_{0 j}|\psi_{0}^{(g)}\rangle\langle\psi_{j}^{(e)}|,
\end{equation}
i.e. only those elements of the optical coherences are different from zero in the first order, which ``catch'' on the dark state. All nonzero elements of $\hat{\tilde{\rho}}_{eg}^{(1)}$ and $\hat{\tilde{\rho}}_{ge}^{(1)}$ are found from the equation
\begin{equation}\label{J_J:eq_rho_gg}
    \frac{\partial}{\partial\,t}\hat{\rho}_{gg}^{(0)} = \frac{i}{\hbar}\left(\hat{V}^{\dagger}\hat{\tilde{\rho}}_{eg}^{(1)} - \hat{\tilde{\rho}}_{ge}^{(1)}\hat{V}\right).
\end{equation}
Substituting (\ref{J_J:V_def_diag}) and (\ref{J_J:rho_gg_0}) to (\ref{J_J:eq_rho_gg}) and multiplying obtained equation on the left by $\langle\psi_{j}^{(g)}|$ and on the right by $|\psi_{0}^{(g)}\rangle$, we have
\begin{equation}\label{J_J:psi_k_e_rho_eg_psi_0_g}
    \langle\psi_{j}^{(e)}|\hat{\tilde{\rho}}_{eg}^{(1)}|\psi_{0}^{(g)}\rangle = - \frac{i\hbar}{\lambda_j^{*}}\langle\psi_{j}^{(g)}|\frac{\partial}{\partial t}\psi_{0}^{(g)}\rangle.
\end{equation}
The time derivative of equation (\ref{J_J:dark_state}) yields the relation
\begin{equation}\label{J_J:diff_V_psi_0}
    \hat{V}|\frac{\partial}{\partial\,t}\psi_{0}^{(g)}\rangle = -\left(\frac{\partial}{\partial\,t}\hat{V}\right)|\psi_{0}^{(g)}\rangle.
\end{equation}
From (\ref{J_J:diff_V_psi_0}) and (\ref{J_J:V_def_diag}) we find
\begin{equation}\label{J_J:psi_k_g_dif_psi_0_g}
    \langle\psi_{j}^{(g)}|\frac{\partial}{\partial\,t}\psi_{0}^{(g)}\rangle = -\frac{1}{\lambda_j}\langle\psi_{j}^{(e)}|\left(\frac{\partial}{\partial t}\hat{V}\right)|\psi_{0}^{(g)}\rangle.
\end{equation}
Substituting (\ref{J_J:psi_k_g_dif_psi_0_g}) into (\ref{J_J:psi_k_e_rho_eg_psi_0_g}), we obtain expression for the nonzero matrix elements of the optical coherence $\hat{\tilde{\rho}}_{eg}^{(1)}$ in basis (\ref{J_J:basis_eigenstates}):
\begin{equation}\label{J_J:rho_eg_t}
    \hat{\tilde{\rho}}_{eg}^{(1)} = i\hbar\sum_{j\neq0}\frac{1}{|\lambda_j|^2}|\psi_{j}^{(e)}\rangle \langle\psi_{j}^{(e)}|\left(\frac{\partial}{\partial\,t}\hat{V}\right)|\psi_{0}^{(g)}\rangle\langle\psi_{0}^{(g)}|.
\end{equation}
The polarization vector, using the formulae (\ref{method:Pt_def_from_micro-param}) and (\ref{J_J:rho_eg_t}), can be expressed as
\begin{equation}\label{J_J:P_vec}
    {\bf\tilde{P}}(t,z) = i\hbar N_a \hat{M}\,\frac{\partial}{\partial\,t}{\bf\tilde{E}},
\end{equation}
where the components of the Hermitian matrix $\hat{M}$ are
\begin{equation}\label{J_J:M}
    M_{q}{}^{q'} = \sum_{j\neq0}\frac{1}{|\lambda_j|^{2}}\langle\psi_{0}^{(g)}|(\hat{d}_{eg}^{1q})^{\dag} |\psi_{j}^{(e)}\rangle\langle\psi_{j}^{(e)}|\hat{d}_{eg}^{1q'} |\psi_{0}^{(g)}\rangle.
\end{equation}
From (\ref{pulses:M e_bot=m e_bot}) and (\ref{J_J:M}) we find
\begin{eqnarray}\label{basic:m_bot_def}
        &&m_{\bot}  = ({\bf e}_{\bot}^{\,*}\cdot \hat{M}{\bf e}_{\bot}) = \nonumber\\
        &&\frac{1}{A^2}\sum_{j\neq0}\frac{1}{|\lambda_j|^2}\langle\psi_{0}^{(g)}|\Big(\frac{\partial}{\partial \varepsilon}\hat{V}\Big)^{\dag} |\psi_{j}^{(e)}\rangle\langle\psi_{j}^{(e)}|\Big(\frac{\partial}{\partial \varepsilon}\hat{V}\Big)|\psi_{0}^{(g)}\rangle.\nonumber\\
\end{eqnarray}
The derivative of (\ref{J_J:dark_state}) with respect to ellipticity angle $\varepsilon$ yields the identity
\begin{equation}\label{J_J:diff_V_psi_0_epsilon}
    \left(\frac{\partial}{\partial \varepsilon}\hat{V}\right)|\psi_{0}^{(g)}\rangle = -\hat{V}|\frac{\partial}{\partial \varepsilon}\psi_{0}^{(g)}\rangle.
\end{equation}
Substituting (\ref{J_J:diff_V_psi_0_epsilon}) into (\ref{basic:m_bot_def}) and taking into consideration the relations
\begin{align}
    \label{J_J:condition_fullness}
        \sum_{j\neq0}|\psi_{j}^{(g)}\rangle\langle\psi_{j}^{(g)}| = \hat{1} - |\psi_{0}^{(g)}\rangle\langle\psi_{0}^{(g)}|,\\
    \label{J_J:real_dark_func}
        \langle\frac{\partial}{\partial\varepsilon}\psi_{0}^{(g)}|\psi_{0}^{(g)}\rangle = \langle\psi_{0}^{(g)} | \frac{\partial}{\partial\varepsilon}\psi_{0}^{(g)}\rangle = 0,
\end{align}
we find
\begin{equation}\label{J_J:m_psi0}
    m_{\bot} = \frac{1}{A^2}\langle\frac{\partial}{\partial \varepsilon}\psi_{0}^{(g)}|\frac{\partial}{\partial \varepsilon}\psi_{0}^{(g)}\rangle.
\end{equation}
Let us express the slowing factor $m_{\bot}$ in the following form
\begin{eqnarray}
        &&m_{\bot} =
        \left.
            \frac{1}{A^2}\langle\frac{\partial}{\partial \varepsilon_1}\psi_{0}^{(g)}({\bf e}_1)|\frac{\partial}{\partial \varepsilon_2}\psi_{0}^{(g)}({\bf e}_2)\rangle
        \right|_{\varepsilon_1 = \varepsilon_2 = \varepsilon}\nonumber\\
        &&=
        \left.
            \frac{1}{A^2}\frac{\partial^2}{\partial \varepsilon_1 \partial \varepsilon_2}\langle\psi_{0}^{(g)}({\bf e}_1)|\psi_{0}^{(g)}({\bf e}_2)\rangle
        \right|_{\varepsilon_1 = \varepsilon_2 = \varepsilon},
\end{eqnarray}
i.e. we first calculate the scalar product for dark states associated with two independent elliptical polarization vectors ${\bf e}_1$  and ${\bf e}_2$, then we put in obtained expression $\varepsilon_1 = \varepsilon_2 = \varepsilon$. Major axis of polarization ellipses are considered parallel (i.e. $\phi_1 = \phi_2 = \phi$), that does not affect the final answer.

\begin{figure}[h!]
    \includegraphics[width=0.9\linewidth]{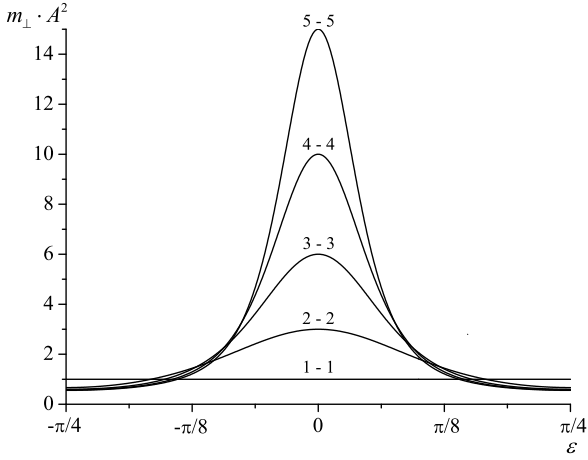}
    \caption{\label{image:m_botDotA_J-J} Slowing factor $m_{\bot}$ as function of ellipticity angle $\varepsilon$ for the transitions $J_g = J \rightarrow J_e = J$ ($J$ is integer).}
\end{figure}

In work  \cite{TTY_PRA_2004} the explicit invariant form of dark states were represented in form of expansion in the Zeeman wave functions $\left|J_g, \mu_g\right\rangle$ has been found:
\begin{equation}\label{J_J:expansion_psi}
    |\Psi^{(nc)}\rangle \equiv |\psi_0^{(g)}\rangle = \sum_{\mu_g}(-1)^{-\mu_g}\Psi_{J_g -\mu_g}^{(nc)}|J_g, \mu_g\rangle.
\end{equation}
For the transition $J_e = J \rightarrow J_g = J$ ($J$ is integer) in \cite{TTY_PRA_2004} it was obtained
\begin{equation}\label{J_J:tensor_Psi_Jg_NC}
    \Psi_{J}^{(nc)}= \mathcal{N}\{{\bf e}\}_{J},
\end{equation}
normalization factor $\mathcal{N}$  is given the formula
\begin{equation}\label{J_J:norm_const_N}
    \mathcal{N} = \left[\frac{J!}{\left(2J - 1\right)!!}\left|{\bf e}\cdot{\bf e}\right|^{J} P_{J}\left(\frac{1}{\left|{\bf e}\cdot{\bf e}\right|}\right)\right]^{-\frac{1}{2}},
\end{equation}
where $P_{L}(x)$ is the Legendre polynomial. Here we have used the notation for tensor product of the identical vectors whose rank is equal to the number of its constituent vectors \cite{Manakov_JETP_1997}
\begin{equation}
    \{{\bf e}\}_{J} = \{...\{\{{\bf e}\otimes{\bf e}\}_2\otimes{\bf e}\}_3...\otimes{\bf e}\}_{J}.
\end{equation}
An expression $\{D_{L1}\otimes D_{L2}\}_{L}$ is the irreducible tensor product of rank $L$ of the irreducible tensors $D_{L1}$ and $D_{L2}$ of ranks $L1$ and $L2$, respectively (for example, see \cite{Varshalovich}).

From expression for wave polarization vector (\ref{pulses:e_def}) we find that scalar product $({\bf e}\cdot{\bf e})$ has the form
\begin{equation}\label{J_J:e_dot_e}
   ({\bf e}\cdot{\bf e}) = \cos{2\varepsilon} \geq 0 \quad \left(-\pi/4\leq\varepsilon\leq\pi/4\right).
\end{equation}
Using the results of \cite{Manakov_JETP_1997}, we find
\begin{eqnarray}\label{J_J:a_dot_b}
        &&(\{{\bf e}_1\}_J,\{{\bf e}_2\}_J) =
        \nonumber\\
        &&\frac{J!}{(2J - 1)!!}\big(({\bf e}_1\cdot{\bf e}_1)({\bf e}_2^{*}\cdot{\bf e}_2^{*})\big)^{\frac{J}{2}} P_J\bigg(\frac{({\bf e}_1\cdot{\bf e}_2^{*})}{\sqrt{({\bf e}_1\cdot{\bf e}_1)({\bf e}_2^{*}\cdot{\bf e}_2^{*})}}\bigg)
        \nonumber\\
        && = \frac{J!}{(2J - 1)!!}\big(\sqrt{\cos{2\varepsilon_1} \cos{2\varepsilon_2}}\big)^J P_J\bigg(\frac{\cos{(\varepsilon_1 - \varepsilon_2)}}{\sqrt{\cos{2\varepsilon_1} \cos{2\varepsilon_2}}}\bigg).
        \nonumber\\
\end{eqnarray}
The final expression for $m_{\bot}$ can be written as
\begin{equation}\label{J_J:m_bot_final}
    m_{\bot} = \frac{1}{A^2}\left[x_{\varepsilon}^2(x_{\varepsilon}^2 - 1)\frac{d^2}{d x_{\varepsilon}^2} + x_{\varepsilon}^3\frac{d}{d x_{\varepsilon}}\right]\ln{P_J(x_{\varepsilon})},
\end{equation}
here, we use the notation
\begin{equation}\label{J_J:x_eps}
    x_{\varepsilon} = \frac{1}{\cos{2\varepsilon}}.
\end{equation}
From the general expression (\ref{J_J:m_bot_final}) for pulses with arbitrary ellipticity a simple form of $m_\bot$ follows for two extreme particular cases of polarization: linear
\begin{equation}\label{J_J:m_bot_linear}
    m_\bot(J, \varepsilon = 0) = \frac{1}{A^2}\frac{J(J+1)}{2}
\end{equation}
and circular
\begin{equation}\label{J_J:m_bot_circular}
    m_\bot(J, \varepsilon = \pm\pi/4) = \frac{1}{A^2}\frac{J}{2J - 1}.
\end{equation}
The dependence of slowing factor $m_{\bot}$ on elipticity angle $\varepsilon$ for some transitions is shown in Fig.~\ref{image:m_botDotA_J-J}.
%

%
%--------------------- SLOWING FACTORS FOR THE J -> J-1 TRANSITIONS ------------------------
\section{\label{Appendix:JtoJ-1}Transition $J_g = J \rightarrow J_e = J -1$ ($J$ is arbitrary)}
Let us consider other transitions $J_g = J \rightarrow J_e = J -1$ ($J$ is arbitrary) with coherent population trapping. For such level structure, according to \cite{Yudin_1989}, by resonant interaction with the elliptically polarized radiation there are two independent systems of the Zeeman sublevels associated light-induced transitions. One of them begins from sublevel  $|J_g, \mu = - J_g\rangle$, and other begins from the $|J_g, \mu = - J_g + 1\rangle$. In this case, there are always two independent dark states $|\psi_{1}^{(nc)}\rangle$ and $|\psi_{2}^{(nc)}\rangle$, which are coherent superposition of the Zeeman sublevels of ground level and satisfy the equations:
\begin{equation}\label{J_J-1:dark states}
    \hat{V}|\psi_{1}^{(nc)}\rangle = 0, \quad  \hat{V}|\psi_{2}^{(nc)}\rangle = 0.
\end{equation}
Zeroth-order approximation for density matrix $\hat{\tilde{\rho}}^{(0)}$ in the expansion (\ref{method:expansion_rho}) corresponds to stationary solution for monochromatic wave and is determined by contribution of each dark state
\begin{equation}\label{J_J-1:rho_0_def}
    \begin{aligned}
        &\hat{\rho}_{gg}^{(0)} = p_{1}|\psi_{1}^{(nc)}\rangle\langle\psi_{1}^{(nc)}| + p_{2}|\psi_{2}^{(nc)}\rangle\langle\psi_{2}^{(nc)}|,\\
        &\hat{\rho}_{ee}^{(0)} = 0, \quad \hat{\rho}_{eg(ge)}^{(0)} = 0,
    \end{aligned}
\end{equation}
where $p_{1} + p_{2} = 1$, $p_{1}$ and $p_{2}$ are the population of dark states. In the first order perturbation theory from equation (\ref{method:rho_1}) taking into account (\ref{J_J-1:rho_0_def}) for optical coherences it follows that are different from zero only elements, which end in one of the dark states
\begin{eqnarray}
\label{J_J-1:rho_eg_def}
    \hat{\tilde{\rho}}_{eg}^{(1)} &=& \sum_{j}\left(\rho_{j1}^{eg}|\psi_{j}^{(e)}\rangle\langle\psi_{1}^{(nc)}| + \rho_{j2}^{eg}|\psi_{j}^{(e)}\rangle\langle\psi_{2}^{(nc)}|\right),\\
\label{J_J-1:rho_ge_def}
    \hat{\tilde{\rho}}_{ge}^{(1)} &=& \sum_{j}\left(\rho_{1j}^{ge}|\psi_{1}^{(nc)}\rangle\langle\psi_{j}^{(e)}| + \rho_{2j}^{ge}|\psi_{2}^{(nc)}\rangle\langle\psi_{j}^{(e)}|\right).
\end{eqnarray}
From (\ref{J_J:V_def_diag}), (\ref{J_J:eq_rho_gg}) and (\ref{J_J-1:rho_0_def}) we find
\begin{equation}\label{J_J-1:rho_eg_j12}
    \rho_{j1(2)}^{eg} = - \frac{i\hbar}{\lambda_{j}^{*}}p_{1(2)} \langle\psi_{j}^{(g)}|\frac{\partial}{\partial t}\psi_{1(2)}^{(nc)}\rangle.
\end{equation}
Using the identity
\begin{equation}\label{J_J-1:psi_k_g_dif_psi_12_nc}
    \langle\psi_{j}^{(g)}|\frac{\partial }{\partial t}\psi_{1(2)}^{(nc)}\rangle = -\frac{1}{\lambda_j}\langle\psi_{j}^{(e)}|\Big(\frac{\partial }{\partial t}\hat{V}\Big)|\psi_{1(2)}^{(nc)}\rangle,
\end{equation}
from  (\ref{J_J-1:rho_eg_def}) and (\ref{J_J-1:rho_eg_j12}) we obtain expression for $\hat{\tilde{\rho}}_{eg}^{(1)}$
\begin{eqnarray}\label{J_J-1:rho_eg}
        \hat{\tilde{\rho}}_{eg}^{(1)} &=& i\hbar\sum_{j \neq 1, 2}\frac{1}{|\lambda_j|^2} \Big[p_{1}|\psi_{j}^{(e)}\rangle\langle\psi_{j}^{(e)}|\Big(\frac{\partial }{\partial t}\hat{V}\Big)|\psi_{1}^{(nc)}\rangle\langle\psi_{1}^{(nc)}|
        \nonumber\\
        &+& p_{2}|\psi_{j}^{(e)}\rangle\langle\psi_{j}^{(e)}|\Big(\frac{\partial }{\partial t}\hat{V}\Big)|\psi_{2}^{(nc)}\rangle\langle\psi_{2}^{(nc)}|\Big].
\end{eqnarray}
Polarization vector of the medium as in the case of $J \rightarrow J$ transitions can be written in the form
\begin{equation}\label{J_J-1:P_vec}
    {\bf\tilde{P}}(t,z) = i\hbar N_a \hat{M}\,\frac{\partial}{\partial\,t}{\bf\tilde{E}},
\end{equation}
where components of the Hermit matrix $\hat{M}$ are given by
\begin{eqnarray}\label{J_J-1:M}
        M_{q}{}^{q'} &=&\sum_{j \neq 1, 2}\frac{1}{|\lambda_j|^2} \Big[p_{1}\langle\psi_{1}^{(nc)}|\big(\hat{d}_{eg}^{1q}\big)^{\dag}|\psi_{j}^{(e)}\rangle \langle\psi_{j}^{(e)}|\hat{d}_{eg}^{1q'}|\psi_{1}^{(nc)}\rangle
        \nonumber\\
        &+& p_{2}\langle\psi_{2}^{(nc)}|\big(\hat{d}_{eg}^{1q}\big)^{\dag}|\psi_{j}^{(e)}\rangle \langle\psi_{j}^{(e)}|\hat{d}_{eg}^{1q'}|\psi_{2}^{(nc)}\rangle\Big].
\end{eqnarray}
For the eigenvalue $m_\bot$ of matrix  $\hat{M}$ we obtain the expression
\begin{equation}\label{J_J-1:m_bot_psi_12_nc}
    m_{\bot} = \frac{1}{A^2}\Big[p_{1} \langle\frac{\partial }{\partial\varepsilon}\psi_{1}^{(nc)} | \frac{\partial}{\partial\varepsilon}\psi_{1}^{(nc)}\rangle + p_{2} \langle\frac{\partial }{\partial\varepsilon}\psi_{2}^{(nc)} | \frac{\partial}{\partial\varepsilon}\psi_{2}^{(nc)}\rangle\Big].
\end{equation}
Derivation of the formula (\ref{J_J-1:m_bot_psi_12_nc}) is similar to one of the formula (\ref{J_J:m_psi0}) with taking into consideration the relations
\begin{align}
    &\langle\psi_{j}^{(e)}|\Big(\frac{\partial }{\partial\varepsilon}\hat{V}\Big)|\psi_{1(2)}^{(nc)}\rangle = - \lambda_j\langle\psi_{j}^{(g)}|\frac{\partial }{\partial\varepsilon}\psi_{1(2)}^{(nc)}\rangle,\\
    &\sum_{j \neq 1, 2}|\psi_{j}^{(g)}\rangle\langle\psi_{j}^{(g)}| = \hat{1} - \sum_{k = 1, 2} |\psi_{k}^{(nc)}\rangle\langle\psi_{k}^{(nc)}|,\\
    &\langle\frac{\partial}{\partial\varepsilon}\psi_{1(2)}^{(nc)}|\psi_{1(2)}^{(nc)}\rangle = \langle\psi_{1(2)}^{(nc)} | \frac{\partial}{\partial\varepsilon}\psi_{1(2)}^{(nc)}\rangle = 0
\end{align}
and orthogonality of the dark states.

\begin{figure}[h!]
    \includegraphics[width=0.95\linewidth]{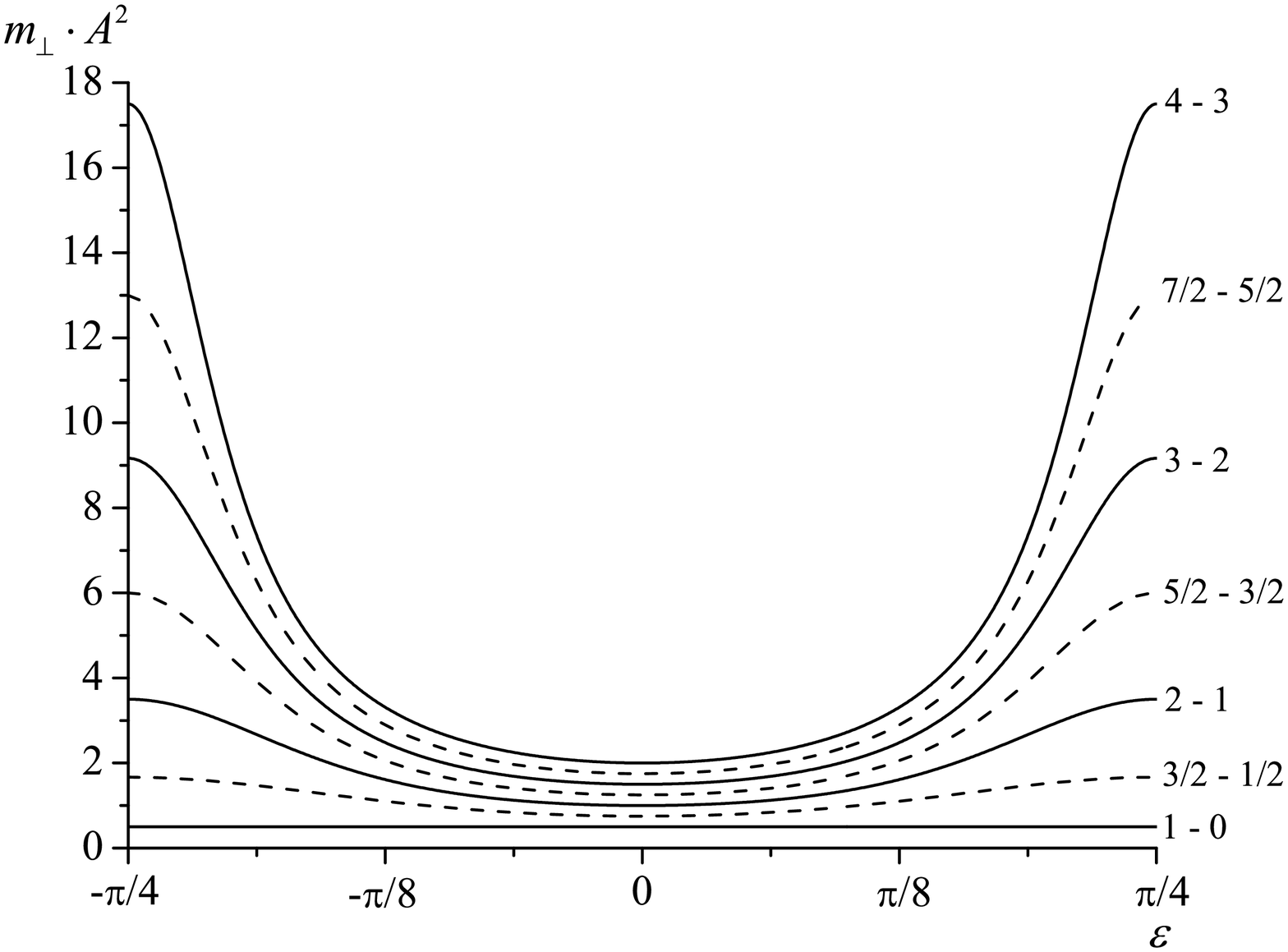}
    \caption{\label{image:m_botDotA_J-Jminus1} Slowing factor $m_{\bot}$ as function of ellipticity angle $\varepsilon$ for the transitions $J_g = J \rightarrow J_e = J - 1$: $J$ is integer (solid line) and $J$ is half-integer (dashed line).}
\end{figure}

For convenience in subsequent calculations we denote as $|\Psi_{+}^{(nc)}\rangle$ the dark state connected with first chain of sub-levels, and as $|\Psi_{-}^{(nc)}\rangle$ the dark state connected with second  $\Lambda$-chain
\begin{equation}\label{J_J-1:new_dark_notations}
    |\Psi_{+}^{(nc)}\rangle = |\psi_{1}^{(nc)}\rangle, \quad |\Psi_{-}^{(nc)}\rangle = |\psi_{2}^{(nc)}\rangle.
\end{equation}
The form of these states has been derived in \cite{TTY_PRA_2004}
\begin{equation}
    |\Psi_{\pm}^{(nc)}\rangle = \sum_{\mu_g}(-1)^{-\mu_g}\Psi^{(\pm)}_{J_g -\mu_g}|J_g, \mu_g\rangle,
\end{equation}
where two linearly independent orthonormal tensors $\Psi^{(+)}_J$ and $\Psi^{(-)}_J$ were expressed in terms of a pair of circular vectors
\begin{equation}\label{J_J-1:tensor_Psi_pm}
    \Psi^{(\pm)}_J = \frac{\{{\bf c}^{(1)}\}_J \pm \{{\bf c}^{(2)}\}_J}{\sqrt{2\left[1 \pm \left(\{{\bf c}^{(1)*}\}_J \cdot \{{\bf c}^{(2)}\}_J\right)\right]}}.
\end{equation}
Vectors ${\bf c}^{(1)}$ and ${\bf c}^{(2)}$ are defined by the field polarization vector the following invariant way
\begin{equation}\label{J_J-1:vec_c_1_2_def}
    {\bf c}^{(1, 2)} = \frac{\left[{\bf e} \times \left[{\bf e} \times {\bf e}^{*}\right]\right] \pm i\left[{\bf e} \times {\bf e}^{*}\right]\sqrt{({\bf e} \cdot {\bf e})}}{\sqrt{\left(1 - |{\bf e} \cdot {\bf e}|^2\right)\left(1 + |{\bf e} \cdot {\bf e}|\right)}}.
\end{equation}
Substituting the expression for the wave polarization vector (\ref{pulses:e_def}) in the formula (\ref{J_J-1:vec_c_1_2_def}) we find
\begin{eqnarray}\label{J_J-1:vec_c_1_2}
        {\bf c}^{(1, 2)} &=& \frac{\sin{2\varepsilon}}{\cos{\varepsilon}\sqrt{2\sin^{2}{2\varepsilon}}}\Big[\sin{\Big(\varepsilon - \frac{\pi}{4}\Big)}e^{i\phi}{\bf e}_{-1}
        \nonumber\\
        &-& \cos{\Big(\varepsilon - \frac{\pi}{4}\Big)}e^{-i\phi}{\bf e}_{+1} \pm \sqrt{\cos{2\varepsilon}}\, {\bf e}_0\Big].
\end{eqnarray}
In the case when one of the vectors ${\bf a}$ or ${\bf b}$ (or both) in the scalar product (\ref{J_J:a_dot_b}) is circular, then it is right the identity
\begin{equation}\label{J_J-1:a_dot_b_circular}
    (\{{\bf a}\}_J , \{{\bf b}\}_J) = (\{{\bf a}\}_J \cdot \{{\bf b}^{*}\}_J) = ({\bf a} \cdot {\bf b}^{*})^J.
\end{equation}
With account for the relation (\ref{J_J-1:a_dot_b_circular}) and expression (\ref{J_J-1:vec_c_1_2}), from (\ref{J_J-1:tensor_Psi_pm}) it follows
\begin{equation}\label{J_J-1:tensor_Psi_pm_1}
    \Psi^{(\pm)}_J = \frac{\{{\bf c}^{(1)}\}_J \pm \{{\bf c}^{(2)}\}_J}{\sqrt{2\left[1 \pm (\tan{\varepsilon})^{2J}\right]}}.
\end{equation}
Differentiate the equation  (\ref{J_J-1:tensor_Psi_pm_1}) with respect to ellipticity angle $\varepsilon$
\begin{eqnarray}\label{J_J-1:def_vareps_tensor_Psi_pm}
    &&\frac{\partial}{\partial \varepsilon}\Psi^{(\pm)}_J = \big[2\big(1 \pm (\tan{\varepsilon})^{2J}\big)\big]^{-\frac{1}{2}} \Big(\frac{\partial}{\partial \varepsilon}\{{\bf c}^{(1)}\}_J
    \nonumber\\
    && \pm \frac{\partial}{\partial \varepsilon}\{{\bf c}^{(2)}\}_J\Big) \mp 2J(\tan{\varepsilon})^{2J-1}(1 + \tan^2{\varepsilon})
    \nonumber\\
    && \times \big[2\big(1 \pm (\tan{\varepsilon})^{2J}\big)\big]^{-\frac{3}{2}}\big(\{{\bf c}^{(1)}\}_J \pm \{{\bf c}^{(2)}\}_J\big)
\end{eqnarray}
and using the method of calculating the scalar products in (\ref{J_J-1:m_bot_psi_12_nc}) described in detail for $J \rightarrow J$  ($J$ is integer) transitions, we obtain
\begin{eqnarray}\label{J_J-1:cdot_nc_pm}
    &&\langle\frac{\partial}{\partial \varepsilon}\Psi_{\pm}^{(nc)}|\frac{\partial}{\partial \varepsilon}\Psi_{\pm}^{(nc)}\rangle =
    \nonumber\\
    &&\left.
    \left(\frac{\partial}{\partial \varepsilon_1} \Psi_J^{(\pm)}({\bf e}_1), \frac{\partial}{\partial \varepsilon_2} \Psi_J^{(\pm)}({\bf e}_2)\right)
    \right|_{\varepsilon_1 = \varepsilon_2 = \varepsilon} =
    \nonumber\\
    && \frac{J}{2}\frac{(1 + y_{\varepsilon}^2)^2}{1 \pm y_{\varepsilon}^{2J}}\left[\frac{1 \mp y_{\varepsilon}^{2J-2}}{1 - y_{\varepsilon}^2} \pm \frac{2Jy_{\varepsilon}^{2J-2}}{1 \pm y_{\varepsilon}^{2J}}\right],
\end{eqnarray}
where the notation is introduced
\begin{equation}\label{J_J-1:y_varepsilon}
    y_{\varepsilon} = \tan{\varepsilon}.
\end{equation}
In deriving of (\ref{J_J-1:cdot_nc_pm}) the following relations were used
\begin{eqnarray}
    &&(\{{\bf c}^{(1)}(\varepsilon_1)\}_J,\{{\bf c}^{(1)}(\varepsilon_2)\}_J) = (\{{\bf c}^{(2)}(\varepsilon_1)\}_J,\{{\bf c}^{(2)}(\varepsilon_2)\}_J)
    \nonumber\\
    &&= \left(\frac{\cos{(\varepsilon_1 - \varepsilon_2)} + \sqrt{\cos{2\varepsilon_1}\cos{2\varepsilon_2}}}{2\cos{\varepsilon_1}\cos{\varepsilon_2}}\right)^J,
\end{eqnarray}
\begin{eqnarray}
    &&(\{{\bf c}^{(1)}(\varepsilon_1)\}_J,\{{\bf c}^{(2)}(\varepsilon_2)\}_J) = (\{{\bf c}^{(2)}(\varepsilon_1)\}_J,\{{\bf c}^{(1)}(\varepsilon_2)\}_J)
    \nonumber\\
    &&= \left(\frac{\cos{(\varepsilon_1 - \varepsilon_2)} - \sqrt{\cos{2\varepsilon_1}\cos{2\varepsilon_2}}}{2\cos{\varepsilon_1}\cos{\varepsilon_2}}\right)^J.
\end{eqnarray}
Thus, the slowing factor $m_{\bot}$ for polarization pulses in case of the transitions $J \rightarrow J-1$ ($J$ is arbitrary) is given by
\begin{eqnarray}\label{J_J-1:m_bot_final}
    &&m_{\bot} = \frac{J(1 + y_{\varepsilon}^2)^2}{2A^2} \bigg[p_1\bigg(\frac{1 - y_{\varepsilon}^{2J-2}}{(1 - y_{\varepsilon}^2)(1 + y_{\varepsilon}^{2J})} + \frac{2Jy_{\varepsilon}^{2J-2}}{(1 + y_{\varepsilon}^{2J})^2}\bigg)
    \nonumber\\
    &&+ p_2\bigg(\frac{1 + y_{\varepsilon}^{2J-2}}{(1 - y_{\varepsilon}^2)(1 - y_{\varepsilon}^{2J})} - \frac{2Jy_{\varepsilon}^{2J-2}}{(1 - y_{\varepsilon}^{2J})^2}\bigg)\bigg].
\end{eqnarray}
From this general expression assuming $p_1 = p_2 = 1/2$ a simple form of $m_\bot$ follows for two extreme particular cases of polarization, linear:
\begin{equation}\label{J_J:m_bot_linear}
    m_\bot(J, \varepsilon = 0) = \frac{1}{A^2}\frac{J}{2}\,,
\end{equation}
and circular:
\begin{equation}\label{J_J:m_bot_circular}
    m_\bot(J, \varepsilon = \pm\pi/4) = \frac{1}{A^2}\frac{(4J-1)(2J-1)}{6}\,.
\end{equation}
The dependence of $m_{\bot}$ on elipticity angle $\varepsilon$ for some transitions is shown in Fig.~\ref{image:m_botDotA_J-Jminus1}. The population of the dark states is assumed equal ($p_1 = p_2 = 1/2$).

%-------------------------- BIGLIOGRAPHY ------------------------------------

\end{document}